\documentclass[aps,preprintnumbers,nofootinbib,superscriptaddress,11pt]{revtex4}
\usepackage[hypertex]{hyperref}
\usepackage[dvipdfmx]{graphicx} 
\usepackage{amsmath,amssymb,amsfonts,bm,cancel} 

\def\a{\alpha}         \def\th{\theta}     \def\m{\mu} \def\n{\nu}      \def\s{\sigma}  \def\t{\tau}       

\def\dg{\dagger}

\newcommand{\lsp}{ \left ( } \newcommand{\rsp}{ \right ) }   \newcommand{\To}{\Rightarrow}  \newcommand{\getsto}{\leftrightarrow} 

\newcommand{\vev}[1]{ \langle {#1} \rangle }
\def\abs#1{\left| #1\right|}

\newcommand{\row}[2]{ \begin{pmatrix}  #1 & #2   \end{pmatrix}  }

\newcommand{\Column}[3]{ \begin{pmatrix} #1 \\ #2 \\ #3 \end{pmatrix} }
\newcommand{\diag}[2]{ \begin{pmatrix}  #1 & 0 \\ 0 & #2 \\   \end{pmatrix}  }
\newcommand{\offdiag}[2]{ \begin{pmatrix} 0 & #1 \\ #2 & 0 \\   \end{pmatrix} }
\newcommand{\Diag}[3]{ \begin{pmatrix} #1 & 0 & 0 \\ 0 & #2 & 0 \\ 0 & 0 & #3 \\\end{pmatrix}}

\usepackage{color}

\begin{document}


\title{\Large \bf Conditions of general $Z_{2}$ symmetry 
and TM$_{1,2}$ mixing \\   for the minimal type-I seesaw mechanism in an arbitrary basis}

\preprint{STUPP-22-255 }
\author{Masaki J. S. Yang}
\email{yang@krishna.th.phy.saitama-u.ac.jp}
\affiliation{Department of Physics, Saitama University, 
Shimo-okubo, Sakura-ku, Saitama, 338-8570, Japan}
\affiliation{Department of Physics, Graduate School of Engineering Science,
Yokohama National University, Yokohama, 240-8501, Japan}


\begin{abstract} 

In this paper, using a formula for the minimal type-I seesaw mechanism by $LDL^{T}$ (or generalized Cholesky) decomposition, 
conditions of general $Z_{2}$-invariance for the neutrino mass matrix $m$ is obtained in an arbitrary basis.
The conditions are found to be 
$(M_{22} a_{i}^{+} - M_{12} b_{i}^{+}) \,  ( M_{22} a_{j}^{-} - M_{12} b_{j}^{-}) = 
- \det M \, b_{i}^{+} \, b_{j}^{-}$ for the $Z_{2}$-symmetric and -antisymmetric part of a Yukawa matrix $Y_{ij}^{\pm} \equiv (Y \pm T Y )_{ij} /2 \equiv  (a_{j}^{\pm}, b_{j}^{\pm})$ and the right-handed neutrino mass matrix $M_{ij}$.
In other words, the symmetric and antisymmetric part of $b_{i}$ must be proportional to those of the quantity 
$\tilde a_{i} \equiv a_{i} - {M_{12} \over M_{22}} b_{i}$. 
They are equivalent to the condition that $m$ is block diagonalized by eigenvectors of the generator $T$. 

These results are applied to three $Z_{2}$ symmetries, the $\m-\t$ symmetry, the TM$_{1}$ mixing, and the magic symmetry which predicts the TM$_{2}$ mixing. 
For the case of TM$_{1,2}$, the symmetry conditions become 
$ M_{22}^{2} \, \tilde {a}_{1}^{\rm TBM} \tilde a_{2}^{\rm TBM} 
 = - \det M \, b_{1}^{\rm TBM}  b_{2}^{\rm TBM}  $ 
 and 
 $ M_{22}^{2} \, \tilde {a}_{1,2}^{\rm TBM} \tilde a_{3}^{\rm TBM} 
 = - \det M \, b_{1,2}^{\rm TBM} b_{3}^{\rm TBM}$
with components $\tilde a_{i}^{\rm TBM}$ and $b_{i}^{\rm TBM}$ in the TBM basis $\bm v_{1,2,3}$. 
In particular, for the TM$_{2}$ mixing, the magic (anti-)symmetric Yukawa matrix with $S_{2} Y = \pm Y$ 
 is phenomenologically excluded because it predicts $m_{2}=0$ or $m_{1}, m_{3} = 0$.
In the case where Yukawa is not (anti-)symmetric, the mass singular values are displayed without a root sign. 

\end{abstract} 

\maketitle

\section{Introduction}

The structure of the lepton mixing matrix often involves a certain $Z_{2}$ symmetry of the neutrino mass matrix $m$.
The bi-maximal mixing \cite{Barger:1998ta} accompanies the $\m-\t$ symmetry \cite{Fukuyama:1997ky,Lam:2001fb,Ma:2001mr,Balaji:2001ex,Koide:2002cj,Kitabayashi:2002jd, Harrison:2002et,Grimus:2003yn,  Koide:2003rx,Ghosal:2003mq,Aizawa:2004qf,Ghosal:2004qb,Koide:2004gj,Mohapatra:2005yu,Kitabayashi:2005fc,Joshipura:2005vy,Haba:2006hc,Xing:2006xa,Ahn:2006nu, Joshipura:2007sf, GomezIzquierdo:2009id, Xing:2010ez,Ge:2010js,He:2011kn, He:2012yt, Xing:2015fdg, He:2015afa, He:2015xha,  Xing:2017cwb, Gomez-Izquierdo:2017rxi, Fukuyama:2017qxb}, 
and 
the trimaximal mixing \cite{Harrison:2002kp,Friedberg:2006it,Bjorken:2005rm, He:2006qd,Grimus:2008tt,Channey:2018cfj,Bao:2021zwu}
does the magic symmetry \cite{Lam:2006wy}.  
The tri-bi-maximal (TBM) mixing \cite{Harrison:2002er}, the combination of these two mixings 
is realized by the Klein symmetry $K_{4} \simeq Z_{2} \times Z_{2}$ \cite{Lam:2006wm,Lam:2007qc,Lam:2008rs}. 

In this paper, using a recently discovered seesaw formula by $LDL^{T}$ decomposition, 
we investigate conditions of a $Z_{2}$ symmetry in the mass of light neutrinos $m$ 
for the minimal type-I seesaw mechanism \cite{Ma:1998zg, King:1998jw, Frampton:2002qc, Xing:2020ald, Guo:2003cc,Barger:2003gt, Mei:2003gn, Chang:2004wy, Guo:2006qa, Kitabayashi:2007bs, He:2009pt, Yang:2011fh, Harigaya:2012bw, Kitabayashi:2016zec, Bambhaniya:2016rbb, Li:2017zmk, Liu:2017frs, Shimizu:2017fgu, Shimizu:2017vwi, Nath:2018hjx, Barreiros:2018bju, Nath:2018xih,  Wang:2019ovr, Zhao:2020bzx} in an arbitrary basis. 
Such formulation can be applied to other generalized $CP$ symmetries (GCP) 
\cite{Ecker:1981wv, Ecker:1983hz, Gronau:1985sp, Ecker:1987qp,Neufeld:1987wa,Ferreira:2009wh,Feruglio:2012cw,Holthausen:2012dk,Ding:2013bpa,Girardi:2013sza,Nishi:2013jqa,Ding:2013hpa,Feruglio:2013hia,Chen:2014wxa,Ding:2014ora,Ding:2014hva,Chen:2014tpa,Li:2015jxa,Turner:2015uta, Rodejohann:2017lre, Penedo:2017vtf,Nath:2018fvw, Yang:2020qsa, Yang:2020goc, Yang:2021smh, Yang:2021xob} and seesaw mechanisms. 

This paper is organized as follows. 
The next section gives a formula by $LDL^{T}$ decomposition for the minimal type-I seesaw mechanism 
and conditions of general $Z_{2}$ symmetry in an arbitrary basis. 
In Sec.~3, we analyze eigensystems of the $Z_{2}$-symmetric $m$ and its applications. 
The final section is devoted to a summary. 

\section{A formula for the minimal type-I seesaw model and conditions of general $Z_{2}$ symmetry in an arbitrary basis}

Here we review a formula by $LDL^{T}$ decomposition \cite{Yang:2021arl, Yang:2022wch}  in the minimal seesaw models \cite{Ma:1998zg, King:1998jw, Frampton:2002qc}. 
The Yukawa matrix of neutrinos $Y$ and the mass matrix of right-handed neutrinos $M$ are defined as follows
\begin{align}
Y = 
\begin{pmatrix}
a_{1} & b_{1} \\
a_{2} & b_{2} \\
a_{3} & b_{3} \\
\end{pmatrix} \, , 
%
~~~ 
M =
\begin{pmatrix}
M_{11} & M_{12} \\ M_{21} & M_{22}
\end{pmatrix}  \, , 
%
\label{1}
\end{align}
where 
$a_{i}, b_{i}, $ and $M_{ij}$ are general complex parameters. 
By setting the vacuum expectation value of the Higgs field to unity, the mass dimension of $Y_{ij}$ becomes one.
This $M$ can be diagonalized by $LDL^{T}$ (or generalized Cholesky) decomposition;
\begin{align}
\tilde M^{-1} = L^{-1} M^{-1} (L^{-1})^{T}
= \diag{{M_{22} \over \det M} }{1\over M_{22}} 
\equiv \diag{1\over \tilde M_{1}}{1\over \tilde M_{2}}\, , 
\label{tildeM}
\end{align}
where $L$ is a lower unitriangular matrix that has all the diagonal entries equal to one
\begin{align}
L = 
\begin{pmatrix}
1 & 0 \\ - {M_{12} \over M_{22}} & 1
\end{pmatrix} \, , 
~~~ 
L^{-1} =
\begin{pmatrix}
1 & 0 \\  {M_{12} \over M_{22}} & 1
\end{pmatrix} \, . 
\end{align}
By redefining the phase of the second right-handed neutrino $\n_{R 2}$, 
 we can choose a basis such that $M_{22}$ is real-positive. 
A further phase transformation of the first right-handed neutrino $\n_{R 1}' = e^{ i \phi} \n_{R 1}$ yields $\det M' = e^{2 i\phi} (M_{11} M_{22} - M_{12}^{2})$. Thus, the phase of $\det M$ can be absorbed and $\tilde M^{-1}$ can be chosen as real-positive. 
If $M$ is strongly hierarchical, the absolute values of these diagonal elements coincide with the first approximation of the physical singular values of $M$.
In other words, it corresponds to an approximate spectral decomposition based on diagonalization by $L$.

Thus, a deformed Yukawa matrix $\tilde Y$ defined as 
\begin{align}
\tilde Y \equiv 
\begin{pmatrix}
 &  \\[-8pt] \tilde {\bm a} \, & \bm b\\[-8pt]  &  \\
\end{pmatrix} \, 
\equiv \lsp \bm a - \bm b {M_{12} \over M_{22}}  \, , \, \bm b \rsp \, 
= Y 
\begin{pmatrix}
1 & 0 \\ - {M_{12} \over M_{22}} & 1
\end{pmatrix}
\equiv Y L \, , 
%
\label{YL}
\end{align}
yields a formula for the neutrino mass matrix $m$ in an arbitrary basis, 
\begin{align}
m = \tilde Y \tilde M^{-1} \tilde Y^{T} &= 
{M_{22} \over \det M}
\begin{pmatrix}
\tilde a_1^2 & \tilde a_1 \tilde a_2 & \tilde a_1 \tilde a_3 \\
\tilde a_1 \tilde a_2 & \tilde a_2^2 & \tilde a_2 \tilde a_3 \\
\tilde a_1 \tilde a_3 & \tilde a_2 \tilde a_3 & \tilde a_3^2 \\
\end{pmatrix}
+ {1\over M_{22} }
\begin{pmatrix}
b_{1}^{2} & b_{1} b_{2} & b_{1} b_{3} \\
b_{1} b_{2} & b_{2}^{2} & b_{2} b_{3} \\
b_{1} b_{3} & b_{2} b_{3} & b_{3}^{2}
\end{pmatrix}  \\
& \equiv  { 1 \over \tilde M_{1}} \tilde {\bm a} \otimes \tilde {\bm a}^{T} + {1\over \tilde M_{2}} \bm b \otimes \bm b^{T} \, . 
\label{formula}
\end{align}
Although it does not correspond to a physical basis for not hierarchical $M$, 
this formula is valid in any basis.

If the two three-dimensional vectors $\tilde {\bm a} \equiv (\tilde a_{i})$ and $\bm b \equiv (b_{i})$ are linearly independent, the rank of $m$ becomes two and a massless mode appears.
The eigenvector belonging to the zero mode of $m \, m^{\dg}$ is proportional to  $\tilde {\bm a}^{*} \times \bm b^{*}$ (the complex conjugation of the cross product). 


\vspace{12pt}

In this paper, we investigate general conditions under which $m$ has a  $Z_{2}$ symmetry, 
using the formula~(\ref{formula}) and similar arguments as in the previous paper \cite{Yang:2022wch}. 
We obtain more general results than the previous analysis of $CP$-symmetry.

When a Lagrangian has a $Z_2$ symmetry, $m$ satisfies $T \, m \, T^{T} = m$ by a matrix $T$.
Such $Z_{2}$ symmetries include $\m$-$\t$ symmetry \cite{Fukuyama:1997ky,Lam:2001fb,Ma:2001mr} and
magic symmetry \cite{Lam:2006wy}, which has been studied extensively.
$T$ is unitary because the symmetry does not change the kinetic term, 
and $T$ is also Hermitian because $T^{2} =1$ leads to $T = T^{-1} = T^{\dg}$.
Then the three eigenvectors of $T$ belonging to the eigenvalues $\pm 1$ form an orthonormal basis.
The vectors $\tilde {\bm a}$ and $\bm b$ can be expanded by the eigenvectors as $\tilde {\bm a} = \tilde {\bm a}^{+} + \tilde {\bm a}^{-}$ and $ {\bm b} = {\bm b}^{+} + {\bm b}^{-}$. 
This means that the vectors are divided into symmetric and antisymmetric parts 
under the transformation;
\begin{align}
T \tilde {\bm a} = \tilde {\bm a}^{+} - \tilde {\bm a}^{-} \, , ~~~
T {\bm b} = {\bm b}^{+} - {\bm b}^{-} \, .
\label{Ttrf}
\end{align}
Note that either eigenspace is one-dimensional because the eigenvalues are $\{+1, +1, -1\}$ or $\{+1, -1, -1\}$ for a nontrivial $T$. 

Since $\det T$ can be changed by redefinitions of fields, 
it does not lose generality by choosing $\det T = -1$.
In this case, the normalized eigenvectors of $T$ consists of $\bm e^{-}$ belonging to the eigenvalue $-1$ and $\bm e^{+1}, \bm e^{+2}$ belonging to $+1$. 
Using the vector $\bm e^{-}$, the generator can be written as $T = 1 - 2 \bm e^{-} \otimes \bm e^{- \, \dg}$.
If $\bm e^{+1 \, \dg} \bm e^{+2} = 0$ holds, 
$T$ is diagonalized by a unitary matrix $U = (\bm e^{-} \, , \bm e^{+1} \, , \bm e^{+2})$;  
\begin{align}
T' = U^{\dg} T U = 
(\bm e^{-} \, , \bm e^{+1}\, , \bm e^{+2})^{\dg} 
(1 - 2 \bm e^{-} \otimes \bm e^{- \, \dg})
(\bm e^{-} \, , \bm e^{+1}\, , \bm e^{+2})
 = \Diag{-1}{1}{1} \, .
\end{align}
In this basis, a $Z_{2}$-symmetric $m$ must have the following form, 
\begin{align}
m' = U^{\dg} m \, U^{*} = 
\begin{pmatrix}
m_{11}' & 0 & 0 \\
0 & m_{22}' & m_{23}' \\
0 & m_{23}' & m_{33}'
\end{pmatrix} \, .
\end{align}
The conditions for $m$ to be such a block diagonal matrix are
\begin{align}
{ 1 \over \tilde M_{1}}  (\bm e^{-} , \tilde {\bm a} ) (\bm e^{+1} , \tilde {\bm a} ) 
& = - { 1 \over \tilde M_{2}}  (\bm e^{-} , {\bm b} ) (\bm e^{+1} , {\bm b} ) \, , \label{20} \\
{ 1 \over \tilde M_{1}} (\bm e^{-} , \tilde {\bm a} ) (\bm e^{+2} , \tilde {\bm a} ) 
& = - { 1 \over \tilde M_{2}}  (\bm e^{-} , {\bm b} ) (\bm e^{+2} , {\bm b} ) \, , \label{21}
\end{align}
where $(\bm u, \bm v) \equiv \bm u^{\dg} \bm v$ is the Hermitian inner product. 
By multiplying $\bm e^{-}$ and $(\bm e^{+ 1,2})^{T}$ from left and right
of Eqs.~(\ref{20}) and~(\ref{21}) respectively, and adding two equations, 
conditions for $m$ to have a $Z_{2}$ symmetry by $T$ 
are summarized as 
\begin{align}
{1\over \tilde M_{1}} \, \tilde {\bm a}^{-} \otimes \tilde {\bm a}^{+ \, T}
 = -  {1\over \tilde M_{2}} \, \bm b^{-} \otimes \bm b^{+ \, T} \, , 
~~~
M_{22}^{2} \, \tilde a^{+}_{i} \, \tilde a^{-}_{j} = -  \det M \, b^{+}_{i} \, b^{-}_{j} \, .
\label{cond3}
\end{align}
Thus,  $\tilde {\bm a}^{+}$ and $\tilde {\bm a}^{-}$ are proportional to $\bm b^{+}$ and $\bm b^{-}$ respectively and their coefficients are determined by Eq.~(\ref{cond3}). 
Otherwise,  two of the four components will be zero vectors.
This kind of alignment also seems to be  necessary for naturalness in the seesaw mechanism \cite{Yang:2021byq}.

If $\tilde a_{i}^{\pm}$ and $b_{i}^{\pm} \neq 0_{i}$ holds, we can find solutions to $\tilde {\bm a}^{\pm}$ for given $\bm b^{\pm}$; 
\begin{align}
(M_{22} \, \tilde {\bm a}^{+} ,\,  M_{22} \, \tilde {\bm a}^{-}) = ( r \sqrt{\det M}\, \bm b^{+} , \, - {1\over r} \sqrt{\det M} \, \bm b^{-}) \, . 
\label{solution}
\end{align}
Here,  
$r$ is a complex constant defined by the non-zero $\tilde a_{i}^{\pm}$ and $b_{j}^{\pm}$,  
\begin{align}
r \equiv {M_{22} \over \sqrt{\det M}} {\tilde a_{j}^{+} \over b_{j}^{+}} \, 
= - {\sqrt{\det M} \over M_{22} } {b_{i}^{-} \over \tilde a_{i}^{-}} \, .
\label{r}
\end{align}
At first glance, the solution~(\ref{solution}) seems to have a degree of freedom of sign $\pm$.
Indeed, it is correct that some $\tilde a_{i}$ and $- \tilde a_{i}$ are solutions to each other. 
However, they are treated as independent solutions 
because all of $b_{i}^{\pm}$ and one of $\tilde a^{\pm}_{i}$ are required as input parameters 
 to determine the solution uniquely. 
This fact is also manifest in representations~(\ref{rep1}) and (\ref{rep2}) by orthogonal matrices, as seen in the Appendix.

There are four trivial cases with each of $\tilde {\bm a}^{\pm}$ and ${\bm b}^{\pm}$ are $\bm 0$, in which the denominator of $r$ or $1/r$ can not be defined 
and $\tilde {a_{i}}^{\pm}$ need not be proportional to $b_{i}^{\pm}$.
In such cases, $\tilde Y$ itself has definite symmetry, 
\begin{align}
T \, \tilde Y = \{ \tilde Y \, ,  \tilde Y \s_{3} \, ,  - \tilde Y \s_{3} \, ,  - \tilde Y \} \, , 
\label{solution2}
\end{align}
where $\s_{3} \equiv {\rm diag} (1, -1) \, .$
For example, if $\tilde Y$ is $T$-symmetric with $\tilde a_{i}^{-} = b_{i}^{-} = 0$,
we obtain expressions for $Y$ and $a$ from Eq.~(\ref{YL}); 
\begin{align}
a_{i}^{+} =  \tilde a_{i}^{+} + {M_{12} \over M_{22}} b_{i}^{+} \, , ~~~ 
a_{i}^{-} =  0 \, . 
\end{align}
That is, $T \, Y = Y$ holds and the original $Y$ also has the $T$ symmetry.
Although the same is true for antisymmetric $\tilde Y$, 
the other two situations are somewhat different.
For example, the case of $T \, \tilde Y = \tilde Y \s_{3}$, {\it i.e.} $\tilde a^{-}_{i} = b^{+}_{i} = 0$, leads to
\begin{align}
a_{i}^{-} =  {M_{12} \over M_{22}} b_{i}^{-} \, , ~~~ 
a_{i}^{+} =  \tilde a_{i}^{+} \, .
\end{align}
Then, while the antisymmetric components of $a_{i}$ and $b_{i}$ must be proportional, 
$a_{i}$ can have independent symmetric components.
These different behavior influences representations of mass values in the analysis of eigensystems.

\section{Analysis of eigensystem and its application}

Due to the $Z_{2}$ symmetry, the mass values and eigenvectors can be formally determined for each solution~(\ref{solution}) and (\ref{solution2}). 
If Eq.~(\ref{cond3}), 
{\it i.e.} $M_{22}^{2} \, \tilde a^{+}_{i} \, \tilde a^{-}_{j} = - \det M \, b^{+}_{i} \, b^{-}_{j}$ is satisfied, the mass matrix $m$ becomes
\begin{align}
m = 
{1 \over \tilde M_{1}} (\tilde {\bm a}^{+} \otimes \tilde {\bm a}^{+ \, T} + \tilde {\bm a}^{-} \otimes \tilde {\bm a}^{- \, T} ) + {1\over \tilde M_{2} } (\bm b^{+} \otimes \bm b^{+ \, T} + \bm b^{-} \otimes \bm b^{- \, T})  \, .
\label{m}
\end{align}
Since the projective components $\tilde {\bm a}^{\pm}$ and $\bm b^{\mp}$ are orthogonal to each other, the sum of respective projections must be rank one unless $\tilde {\bm a}^{\pm}$ and $\bm b^{\pm}$ (or $\bm b^{\mp}$) are zero vectors.
Thus, each of projections must be proportional to the others, as $\tilde {\bm a}^{\pm} \propto \bm b^{\pm}$. The behavior of the solutions can be classified into the following three cases.

\paragraph{$T \, \tilde Y = \pm \tilde Y \s_{3}$.}

First, if each of $\tilde {\bm a}^{\mp}$ and ${\bm b}^{\pm}$ is $\bm 0$ 
in the trivial solutions~(\ref{solution2}), the mass matrix is
\begin{align}
m = 
{1 \over \tilde M_{1}} \tilde {\bm a}^{\pm} \otimes \tilde {\bm a}^{\pm \, T} + {1 \over \tilde M_{2}} \bm b^{\mp} \otimes \bm b^{\mp \, T}  \, . 
\label{38}
\end{align}
The Hermitian matrix $m \, m^{\dg} $ becomes
\begin{align}
m \, m^{\dg} =  
\left| {\tilde {\bm a}^{\pm} \over \tilde M_{1}} \right |^{2} 
\tilde {\bm a}^{\pm} \otimes \tilde {\bm a}^{\pm \, \dg} 
+ \left | {\bm b^{\mp} \over \tilde M_{2}} \right |^{2} 
\bm b^{\mp} \otimes \bm b^{\mp \, \dg} \, , 
\end{align}
where $|\bm v|^{2} \equiv \sum_{i = 1}^{3} v_{i}^{*} v_{i} .$

The eigenvectors of $m \, m^{\dg}$ are $\{ (\tilde {\bm a}^{\pm} \times \bm b^{\mp})^{*}, \tilde {\bm a}^{\pm} , \bm b^{\mp} \}$, 
and the corresponding mass singular values $m_{i}$ are 
\begin{align}
m_{i} = \{ 0\, , { \abs{\tilde {\bm a}^{\pm}}^{2} \over  | \tilde M_{1} |} \, , {|\bm b^{\mp}|^{2} \over |\tilde M_{2}|}  \} \, .
\label{eigen1}
\end{align}
In this situation, a matrix $S = 1 - 2 \bm v^{+} \otimes \bm v^{+ \, T}$ 
defined by the remaining $\bm v^{+} = \tilde {\bm a}^{+}$ or $\bm b^{+}$
generates another $Z_{2}$ symmetry of $m$ \cite{Lam:2006wm}. 

\paragraph{$T \, \tilde Y = \pm \tilde Y$.}

When $\tilde Y$ and $Y$ are $T$-symmetric or antisymmetric in Eq.~(\ref{solution2}), 
the mass matrix is
\begin{align}
m = 
{1 \over \tilde M_{1}} \tilde {\bm a}^{\pm} \otimes \tilde {\bm a}^{\pm \, T} + {1 \over \tilde M_{2}} \bm b^{\pm} \otimes \bm b^{\pm \, T}  \, . 
\label{40}
\end{align}
Since the eigenvalues of nontrivial $T$ are $(+1, +1, -1)$ or $(-1, -1,+1)$, the solution of $TY= (\det T) Y$ is phenomenologically excluded because the rank of $m$ is one.
In the other solution, 
although the eigenvectors of $m \, m^{\dg}$ are $\{ (\tilde {\bm a}^{\pm} \times \bm b^{\pm})^{*}, \tilde {\bm a}^{\pm} , \bm b^{\pm} \}$, 
there is no guarantee that the two vectors $\tilde {\bm a}^{\pm}$ and $\bm b^{\pm}$ are orthogonal. 
Thus the general representation of mass singular values becomes complicated expressions as displayed in Ref.~\cite{Fujihara:2005pv}.
This is because that $\tilde {\bm a}$ and $\bm b$ belong to the same eigenvalue of $T$.
If we can specify another $Z_{2}$ symmetry by $S$, 
$\tilde {\bm a}$ and $\bm b$ can be decomposed into projections with different eigenvalues of $S$, so that mass singular values can be determined as in the previous case.

\paragraph{Nontrivial solutions.}

In other general situations, 
by substituting the solution~(\ref{solution}) into Eq.~(\ref{m}), 
the mass matrix $m$ is 
\begin{align}
m = 
 {1\over M_{22} } [(1 + r^{2} ) \bm b^{+} \otimes \bm b^{+ \, T} + (1 + {1\over r^{2}}) \bm b^{-} \otimes \bm b^{- \, T} ] \, .
 \label{m3}
\end{align}
The eigenvectors of $m \, m^{\dg}$ are $\{(\bm b^{+} \times \bm b^{-})^{*}, \bm b^{+} , \bm b^{-} \}$ and the mass singular values are
\begin{align}
m_{i} = \{ 0\, , \left | {1+r^{2} \over M_{22}} \right |  |\bm b^{+}|^{2}  \, , 
\left | {1 + {1\over r^{2}} \over M_{22}} \right |  |\bm b^{-}|^{2} \} \, .
\label{eigen2}
\end{align}
There is another $Z_{2}$ symmetry generated by either $S^{\pm} = 1 - 2 \bm b^{\pm} \otimes \bm b^{\pm \, T}$. 
In the following subsections, the results obtained above will be applied to three specific symmetries, the $\m-\t$ symmetry, the TM$_{1}$ mixing, and the TM$_{2}$ mixing predicted from the magic symmetry.

\subsection{Bi-maximal mixing and $\m - \t$ symmetry}

Although the exact $\m - \t$ symmetry \cite{Fukuyama:1997ky} that predicts $\th_{13} = 0$ has now been excluded, let us consider this as a simple example for practice.
First, the basis of the TBM mixing is given by 
\begin{align}
\bm v_{1} = {1\over \sqrt 6} (2, -1, -1)^{T} \, , ~~ 
\bm v_{2} = {1\over \sqrt 3} (1,1,1)^{T} \, , ~~
\bm v_{3} = {1\over \sqrt 2}(0,1,-1)^{T} \, . 
\end{align}
Expanding the vectors $\tilde {\bm a}$ and $\bm b$ in the TBM basis yields
\begin{align}
\tilde {\bm a} & = \bm v_{1}(\bm v_{1} , \tilde {\bm a}) + \bm v_{2} (\bm v_{2}  , \tilde {\bm a}) + \bm v_{3} (\bm v_{3} , \tilde {\bm a} ) \equiv \sum_{i=1}^{3} \tilde a^{\rm TBM}_{i} \bm v_{i} \, ,  \label{expand1} \\
\bm b & = \bm v_{1} (\bm v_{1} , \bm b) + \bm v_{2} (\bm v_{2} , \bm b) + \bm v_{3} (\bm v_{3} , \bm b) \equiv \sum_{i=1}^{3} b^{\rm TBM}_{i} \bm v_{i} \, , 
\label{expand2}
\end{align}
where 
$\tilde a^{\rm TBM}_{i}$ and $b^{\rm TBM}_{i}$ are components in this basis. 

The $\m-\t$ symmetry is defined as\footnote{There is also a definition of $T$ by $v_{3}' = {1 \over \sqrt 2} (0 , 1, 1)$, which is equivalent under the phase transformation $v_{3}' = v_{3}$ diag\,$(1,1,-1)$. }
\begin{align}
T \, m \, T = m \, , ~~~ 
T = 
\begin{pmatrix}
1 & 0 & 0 \\
0 & 0 & 1 \\
0 & 1 & 0 
\end{pmatrix} 
= 1 - 2 \bm v_{3} \otimes \bm v_{3}^{T} \, , 
\label{m-t}
\end{align}
The eigensystem of $T$ is 
\begin{align}
T \bm v_{3} = - \bm v_{3} \, , ~~ T \bm v_{1,2} = + \bm v_{1,2} \, .
\end{align}
Since the third eigenvector of $m \, m ^{\dg}$ is fixed to $v_{3}$, $m_{3}$ can be formally determined as
\begin{align}
m_{3} & = 
{ (\tilde a^{\rm TBM}_{3} )^{2} \over \tilde M_{1}} + {(b^{\rm TBM}_{3})^{2}  \over \tilde M_{2}} 
= 
{M_{22} \over \det M} {(\tilde a_{2} - \tilde a_{3})^{2} \over 2} + {1\over M_{22} } {(b_{2} - b_{3})^{2}\over 2}
 \, .
\end{align}
This expression generally includes complex phases. However, 
 the absolute value $|m_{3}|$ must coincide with the singular value 
because $\bm v_{3}$ belongs to one-dimensional eigenspace of $T$~(\ref{m-t}).

The symmetry condition~(\ref{cond3}) becomes
\begin{align}
M_{22}^{2} \, \tilde {a}_{3}^{\rm TBM} (\tilde a_{1}^{\rm TBM}  \bm v_{1}  + \tilde a_{2}^{\rm TBM} \bm v_{2}  )
 = - \det M \, b_{3}^{\rm TBM} ( b_{1}^{\rm TBM} \bm v_{1} + b_{2}^{\rm TBM} \bm v_{2}) \, . 
\end{align}
Since $\bm v_{1, 2}$ are orthogonal, the equal sign holds for each component;
\begin{align}
M_{22}^{2} \, \tilde {a}_{1}^{\rm TBM} \tilde a_{3}^{\rm TBM} 
 = - \det M \, b_{1}^{\rm TBM}  b_{3}^{\rm TBM}  \, , ~~~ 
 M_{22}^{2} \, \tilde {a}_{2}^{\rm TBM} \tilde a_{3}^{\rm TBM} 
 = - \det M \, b_{2}^{\rm TBM}  b_{3}^{\rm TBM}  \, . 
  \label{symcondmt}
\end{align}
This is equivalent to the block diagonalization conditions~(\ref{20}) and (\ref{21}) in the TBM basis.

Let us examine the three types of solutions analyzed above. 
For the trivial solution~(\ref{38}) with $T \, \tilde Y = \pm \tilde Y \s_{3}$, 
 the deformed Yukawa $\tilde Y$ is 
\begin{align}
\tilde Y = 
\begin{pmatrix}
\tilde a_{1} & 0 \\
\tilde a_{2} & b_{2} \\
\tilde a_{2} & - b_{2} \\
\end{pmatrix}
~~ {\rm or} ~~ 
\begin{pmatrix}
0 & b_{1} \\
\tilde a_{2} & b_{2} \\
- \tilde a_{2} & b_{2} \\
\end{pmatrix} \, . 
\end{align}
The mass matrix $m$ is
\begin{align}
m = {M_{22} \over \det M} 
\begin{pmatrix}
\tilde a_{1}^{2} & \tilde a_{1} \tilde a_{2} & \tilde a_{1} \tilde a_{2} \\
\tilde a_{1} \tilde a_{2} & \tilde a_{2}^{2} & \tilde a_{2}^{2} \\
\tilde a_{1} \tilde a_{2} & \tilde a_{2}^{2} & \tilde a_{2}^{2} 
\end{pmatrix}
 + {1\over M_{22} } 
\begin{pmatrix}
0 & 0 & 0 \\
0 & b_{2}^{2} & - b_{2}^{2} \\
0 & - b_{2}^{2} & b_{2}^{2}
\end{pmatrix}
~~ {\rm or} ~~ (\tilde a_{i} \getsto b_{i}) \, . 
\end{align}
The eigenvectors are $\{ (\tilde {\bm a}^{\pm} \times \bm b^{\mp})^{*}, \tilde {\bm a}^{\pm} , \bm b^{\mp}  \}$ and corresponding singular values $m_{i}$ are 
\begin{align}
m_{1,2} 
&= \{ 0\, ,  {\abs{\tilde a_{1}}^{2} + 2 \abs {\tilde a_{2}}^{2} \over |\tilde M_{1}|}  \}  \, ,
 ~~  m_{3} = \abs{ 2 b_{2}^{2} \over \tilde  M_{2} } \, , \\
 {\rm or} ~~ m_{1,2} & = 
 \{ 0 \, , {\abs {b_{1}}^{2} + 2 \abs {b_{2}}^{2} \over | \tilde M_{2} | }  \} \, , 
 ~~ m_{3} = \abs {  2 \tilde a_{2}^{2} \over \tilde M_{1} } \, .
\end{align}
Since a finite $m_{3}$ is predicted, this is a normal hierarchy (NH) like solution with $\th_{13} = 0$. 
Such Yukawa matrices would be easily realized by flavons with vacuum expectation values  $\vev{\phi_{1}} \propto (-2 , \, 1,\, 1)\, , \vev{\phi_{2}} \propto (1, \, 1,\, 1)\, , \vev{\phi_{3}} \propto (0, \, 1, \, -1)$ \cite{Shimizu:2017fgu}. 

Second, for the other trivial solution~(\ref{40}) in which $\tilde {\bm a}$ and $\bm b$ have the same eigenvalues,  $Y$ and $\tilde Y$ itself has $\m - \t$ (anti-)symmetry $T \, Y = \pm Y$.
\begin{align}
\tilde Y = 
\begin{pmatrix}
\tilde a_{1} & b_{1} \\
\tilde a_{2} & b_{2} \\
\tilde a_{2} & b_{2} \\
\end{pmatrix}
~~ {\rm or} ~~ 
\begin{pmatrix}
0 & 0 \\
\tilde a_{2} & b_{2} \\
- \tilde a_{2} & - b_{2} \\
\end{pmatrix} \, , 
\end{align}
The latter is excluded because $Y$ is also rank one. 
Since the eigenvector $\bm v_{3} = {1 \over \sqrt 2}(0, 1, -1)$ belongs to the zero eigenvalue, 
this solution is inverted hierarchy (IH) like.
In this case, representations of two mass values would be cumbersome.
However, it can be simpler by combining it with another $Z_{2}$ symmetry that accompanies TM$_{1,2}$ mixing described in the next subsections.

Finally, the non-trivial solutions are explored. 
A solution with $\tilde a_{1} = b_{1} = 0$ has no phenomenological interest because the first row and column of $m$ are the zero vector.  
Thus,  $\tilde a_{1}$ and $b_{1}$ are set to have nonzero values and 
the parameter $r$~(\ref{r}) is determined to be $r_{3}= {M_{22} \over \sqrt{\det M}} {\tilde a_{1} \over b_{1}}$.
From Eq.~(\ref{solution}) (or Eq.~(\ref{symcondmt})) the following constraints are obtained; 
The symmetric part is
\begin{align}
\tilde {\bm a}^{+} = 
\Column{\tilde a_{1}}{(\tilde a_{2} + \tilde a_{3})/2}{(\tilde a_{2} + \tilde a_{3})/2}
= {\tilde a_{1} \over b_{1}} \bm b^{+} 
= {\tilde a_{1} \over b_{1}}
\Column{b_{1}}{(b_{2} + b_{3})/2}{(b_{2} + b_{3})/2} \, , ~~~
\end{align}
and the antisymmetric part is
\begin{align}
\tilde {\bm a}^{-} = 
\Column{0}{(\tilde a_{2} - \tilde a_{3})/2}{-(\tilde a_{2} - \tilde a_{3})/2} 
= - {\tilde M_{1} \over \tilde M_{2}} {b_{1} \over \tilde a_{1} } \bm b^{-}
= - {\tilde M_{1} \over \tilde M_{2}} {b_{1} \over \tilde a_{1} }
\Column{0}{(b_{2}- b_{3})/2}{-(b_{2}-b_{3})/2} \, .
\end{align}

Since there are two independent conditions, 
$\tilde a_{2,3}$ are determined for given $\tilde a_{1}$ and $b_{1,2,3}$;
\begin{align}
\tilde a_{2} & =  {\tilde a_{1} \over b_{1}} {b_{2} + b_{3} \over 2} - {\tilde M_{1} \over \tilde M_{2}} {b_{1} \over \tilde a_{1} } {b_{2}- b_{3} \over 2}\, , \\
\tilde a_{3} & =  {\tilde a_{1} \over b_{1}} {b_{2} + b_{3} \over 2} + {\tilde M_{1} \over \tilde M_{2}} {b_{1} \over \tilde a_{1} } {b_{2}- b_{3} \over 2}\,  .
\end{align}
The original Yukawa matrix $Y$ is obtained by the inverse ``rotation'' $L^{-1}$. 
Since $Y$ and $\tilde Y$ have both symmetric and antisymmetric components, they do not have definite symmetry.

The mass matrix~(\ref{m3}) is 
\begin{align}
m 
=  {(1 + r_{3}^{2} )  \over M_{22} } 
\begin{pmatrix}
b_{1}^{2} & b_{1} b_{+} & b_{1} b_{+} \\
b_{1} b_{+} & b_{+}^{2} & b_{+}^{2} \\
b_{1} b_{+} & b_{+}^{2} & b_{+}^{2}
\end{pmatrix}
 + {(1 + {1\over r_{3}^{2}}) \over M_{22}}
\begin{pmatrix}
0 & 0 & 0 \\
0 & b_{-}^{2} & - b_{-}^{2} \\
0 & - b_{-}^{2} & b_{-}^{2}
\end{pmatrix}  \, ,  
\end{align}
where $b_{\pm} \equiv ( b_{2} \pm b_{3}) / 2$.
The eigenvectors are $\{(\bm b^{+} \times \bm b^{-})^{*}, \bm b^{+} , \bm b^{-} \}$ and 
the singular values are
\begin{align}
m_{1,2} = \{ 0\, ,  \abs {1+r_{3}^{2} \over M_{22}} (\abs {b_{1}}^{2} + {\abs {b_{2} + b_{3}}^{2} \over 2}) \} \, , ~~~  m_{3} = \abs {1 + {1\over r_{3}^{2}} \over M_{22}} { |b_{2} - b_{3}|^{2} \over 2} \, . 
\end{align}
Since the $v_{3}$ direction has a nonzero singular value, 
this is also an NH-like solution.

\subsection{TM$_{1}$ mixing}

Recent observations of the non-zero $\th_{13}$ stimulate studies of mixing matrices called TM$_{1,2}$ \cite{Albright:2008rp,Albright:2010ap,He:2011gb,Luhn:2013lkn,Li:2013jya,King:2019vhv,Krishnan:2020xeq};
\begin{align}
U_{\rm TM1} = U_{\rm TBM} U_{23} \, , ~~~
U_{\rm TM2} = U_{\rm TBM} U_{13} \, , ~~~
\end{align}
where 
\begin{align}
U_{\rm TBM} 
= 
\begin{pmatrix}
 \sqrt{\frac{2}{3}} & \frac{1}{\sqrt{3}} & 0 \\
 -\frac{1}{\sqrt{6}} &\frac{1}{\sqrt{3}} & \frac{1}{\sqrt{2}} \\
 -\frac{1}{\sqrt{6}} & \frac{1}{\sqrt{3}} & -\frac{1}{\sqrt{2}} \\
\end{pmatrix} , 
~~~
U_{23} = 
\begin{pmatrix}
1 & 0 & 0 \\
0 & c_{\th} & s_{\th} e^{- i\phi} \\
0 & - s_{\th} e^{i \phi} & c_{\th}
\end{pmatrix} ,
~~~ 
U_{13} =
\begin{pmatrix}
c_{\th} & 0 & s_{\th} e^{- i \phi} \\
0 & 1 & 0 \\
- s_{\th} e^{i \phi} & 0 & c_{\th}
\end{pmatrix} \, , 
\end{align}
with $c_{\th} \equiv \cos \th , \, s_{\th} \equiv \sin \th$. 
The absolute values of $\sin \th_{13}$ are
\begin{align}
|\sin \th _{13}^{\rm TM_{1}} | = \sin \th / \sqrt 3 \, ,  ~~~ 
|\sin \th _{13}^{\rm TM_{2}} | = \sqrt 2 \sin \th / \sqrt {3} \, . 
\end{align}
The following formalism is similar to 
Ref.~\cite{Shimizu:2017fgu} that gives a detailed analysis of TM$_{1}$ and TM$_{2}$ mixing in the minimal seesaw model. 
New points in this paper are that it is presented in an arbitrary basis and the existence of non-trivial solutions.

The matrix $m$ that predicts TM$_{1}$ has a $Z_{2}$ symmetry by the following $S_{1}$; 
\begin{align}
S_{1} \, m \, S_{1} = m \, , ~~ 
S_{1} = 1 - 2 \bm v_{1} \otimes \bm v_{1}^{T} =   
{1\over 3}
\begin{pmatrix}
-1 & 2 & 2 \\
2 & 2 & -1 \\
2 & -1 & 2 \\
\end{pmatrix} \, .
\label{S1}
\end{align}
The eigensystem of $S_{1}$ are 
\begin{align}
S_{1} \bm v_{1} = - \bm v_{1} \, , ~~ S_{1} \bm v_{2,3} = + \bm v_{2,3} \, .
\end{align}
From this, the symmetry conditions~(\ref{20}) and (\ref{21}) are
\begin{align}
M_{22}^{2} \, \tilde {a}_{1}^{\rm TBM} \tilde a_{2,3}^{\rm TBM} 
 = - \det M \, b_{1}^{\rm TBM}  b_{2,3}^{\rm TBM}  \, . 
 \label{symcondTM1}
\end{align}

By transformation to the TBM basis with $U_{\rm TBM} = (\bm v_{1} \, , \bm v_{2} \, , \bm v_{3})$, the mass matrix $m$ with the symmetry conditions is 
\begin{align}
m^{\rm TBM} &\equiv  U^{T}_{\rm TBM} m \, U_{\rm TBM} 
= (\bm v_{1} \, , \bm v_{2} \, , \bm v_{3})^{T}
\lsp { 1 \over \tilde M_{1}} \tilde {\bm a} \otimes \tilde {\bm a}^{T} + {1\over \tilde M_{2}} \bm b \otimes \bm b^{T} \rsp (\bm v_{1} \, , \bm v_{2} \, , \bm v_{3}) \, , \\
& = { 1 \over \tilde M_{1}}
\begin{pmatrix}
( \tilde a_{1}^{\rm TBM})^{2} & 0 & 0 \\
 0 &  (\tilde a_{2}^{\rm TBM})^{2} &  \tilde a_{2}^{\rm TBM}  \tilde a_{3}^{\rm TBM} \\
 0 &  \tilde a_{2}^{\rm TBM}  \tilde a_{3}^{\rm TBM} & (\tilde a_{3}^{\rm TBM} )^{2}
\end{pmatrix}
+  { 1 \over \tilde M_{2}}
\begin{pmatrix}
( b_{1}^{\rm TBM})^{2} & 0 & 0 \\
 0 & ( b_{2}^{\rm TBM})^{2} & b_{2}^{\rm TBM}  b_{3}^{\rm TBM} \\
 0 &  b_{2}^{\rm TBM}  b_{3}^{\rm TBM} & (b_{3}^{\rm TBM})^{2} 
\end{pmatrix} \, .
\end{align}
This $m$ is indeed symmetric under $S_{1}^{\rm TBM} = {\rm diag} (-1 \, , 1 \, , 1)$ in this basis. 
From this, $\th_{13}$ and $m_{1}$ (with a complex phase) are formally obtained as
\begin{align}
|\sin \th_{13}^{\rm TM_{1}}| = \sin \th_{\rm TM_{1}} / \sqrt 3 \, , ~
\tan 2 \th_{\rm TM_{1}} = 
\abs{ {2  \tilde a_{2}^{\rm TBM}  \tilde a_{3}^{\rm TBM} \over \tilde M_{1}}
+ {2 b_{2}^{\rm TBM}  b_{3}^{\rm TBM} \over \tilde M_{2}}
\over 
 \abs{ {(\tilde a_{3}^{\rm TBM} )^{2} \over \tilde M_{1}} + {(b_{3}^{\rm TBM})^{2} \over \tilde M_{2}}} -
   \abs{  {(\tilde a_{2}^{\rm TBM} )^{2} \over \tilde M_{1}} + { (b_{2}^{\rm TBM})^{2} \over \tilde M_{2}} } } \, , 
\end{align}
\begin{align}
m_{1} & = {M_{22} \over \det M} (\tilde {a}_{1}^{\rm TBM} )^{2} + {1\over M_{22} } (b_{1}^{\rm TBM})^{2} 
= {M_{22} \over 6 \det M} (\tilde 2 a_{1} - \tilde a_{2} - \tilde a_{3})^{2} + {1\over 6 M_{22} } (2 b_{1} - b_{2} - b_{3})^{2} \, .
\end{align}

Hereafter mass matrices will be omitted because of their complexity, 
and we will only focus on forms of Yukawa matrices and the mass singular values. 
There are four possibilities for the trivial solutions of TM$_{1}$; 
\begin{align}
S_{1} \tilde Y = \pm \tilde Y ~ \To ~ 
\tilde Y & =
 (x_{2} \bm v_{2} + x_{3} \bm v_{3} , y_{2} \bm v_{2} + y_{3} \bm v_{3} )  
 ~{\rm or}~
 (x_{1} \bm v_{1}, y_{1} \bm v_{1})
\\ & = 
\begin{pmatrix}
{\tilde a_{2} + \tilde a_{3} \over 2} &  {b_{2} + b_{3} \over 2} \\
\tilde a_{2} & b_{2}\\
\tilde a_{3} & b_{3} \\
\end{pmatrix}
~ {\rm or} ~
\begin{pmatrix}
 \tilde a_{1} & b_{1} \\
- \tilde a_{1}/2 & - b_{1}/2  \\
- \tilde a_{1}/2 & - b_{1}/2 \\
\end{pmatrix}
\, , \\
S_{1} \tilde Y = \pm \tilde Y \s_{3} ~ \To ~ 
\tilde Y & = (x_{2} \bm v_{2} + x_{3} \bm v_{3}, y_{1} \bm v_{1} )  ~ {\rm or}~ (x_{1} \bm v_{1}, y_{2} \bm v_{2} + y_{3} \bm v_{3})  \\
& =
\begin{pmatrix}
\tilde a_{1} & { b_{2} + b_{3} \over 2} \\
- \tilde a_{1}/2 & b_{2} \\
- \tilde a_{1}/2 & b_{3} \\
\end{pmatrix} 
~ {\rm or} ~
\begin{pmatrix}
{ \tilde a_{2} + \tilde a_{3} \over 2}& b_{1} \\
\tilde a_{2} & - b_{1}/2 \\
\tilde a_{3} & -b_{1}/2 \\
\end{pmatrix} \, , 
\end{align}
where $x_{1,2,3}$ and $y_{1,2,3}$ are complex coefficients.
Although $S_{1} \tilde Y = \tilde Y$ leads to $m_{1} = 0$ and a NH solution, mass singular values cannot be displayed without solving a complicated quadratic equation. 
A solution with $S_{1} \tilde Y = - \tilde Y$ is excluded because it is rank one.

Solutions $S_{1} \tilde Y = \pm \tilde Y \s_{3}$ lead  to finite $m_{1}$ and IH. 
The mass singular values are
\begin{align}
m_{1} &= \abs{{1\over M_{22}} {3 \over 2} b_{1}^{2}} \, , ~~
m_{2,3} = \{ 0\, , \abs {M_{22} \over \det M} (\abs{\tilde a_{2}}^{2} + \abs{ {\tilde a_{2} + \tilde a_{3}} \over 2}^{2} + \abs{\tilde a_{3}}^{2}) \} \, , \\
~~ {\rm or} ~~ m_{1} & = \abs{{M_{22} \over \det M} {3\over 2} \tilde a_{1}^{2}} \, ,  ~~ 
m_{2,3} = \{ 0\, , \abs {1\over M_{22}} ( \abs{b_{2}}^{2} + \abs{b_{2} + b_{3} \over 2}^{2} + \abs{b_{3}}^{2}) \}  \, . 
\end{align}

Next, nontrivial solutions are investigated. 
In order for the two singlular values~(\ref{eigen2}) to be non-zero, $\tilde {\bm a}$ and $\bm b$ must have $\bm v_{1}$ components.
Thus we can set $(\bm v_{1}, \tilde {\bm a}) \neq 0$ and $(\bm v_{1}, \bm b) \neq 0$.
The parameter  $r$~(\ref{r}) can be determined as 
\begin{align}
r_{1} 
= - {\sqrt{\det M} \over M_{22} } {b_{1}^{\rm TBM}  \over \tilde {a}_{1}^{\rm TBM} }  \, .
\end{align}
By Eq.~(\ref{symcondTM1}), $\tilde a_{2,3}$ can be determined from the other components.
Explicitly, 
\begin{align}
{\tilde a_{1} + \tilde a_{2} + \tilde a_{3} }
&= - {\det M \over M_{22}^{2} } { 2 b_{1} - b_{2} - b_{3} \over  2 \tilde a_{1} - \tilde a_{2} - \tilde a_{3} } { (b_{1} + b_{2} + b_{3}) }
\, , \\
{ \tilde a_{2} -  \tilde a_{3} } 
&= - {\det M \over M_{22}^{2} } { 2 b_{1} - b_{2} - b_{3} \over 2 \tilde a_{1} - \tilde a_{2} - \tilde a_{3} } 
{ (b_{2} -   b_{3}) }
\, .
\end{align}
The mass singular values are obtained as 
\begin{align}
m_{i} &= \{ 0\, ,  \abs{1+r_{1}^{2} \over M_{22}} ( \abs {b_{2}^{\rm TBM} }^{2} + \abs{b_{3}^{\rm TBM} }^{2}) \, , \abs{1 + {1\over r_{1}^{2}} \over M_{22} } \abs{b_{1}^{\rm TBM}}^{2} \}  \\ 
 &= \{ 0\, ,  \abs{1+r_{1}^{2} \over M_{22}} [ {\abs {b_{1}+ b_{2} + b_{3}}^{2} \over 3}+ {\abs { b_{2} -  b_{3}}^{2} \over 2}]  \, , \abs{1 + {1\over r_{1}^{2}} \over M_{22}} { \abs{2 b_{1} - b_{2} - b_{3}}^{2} \over 6} \} \, . 
\end{align}
%

\subsection{TM$_{2}$ mixing and magic symmetry}

Similarly, a mass matrix predicting TM$_{2}$ has a $Z_{2}$ symmetry generated by the following $S_{2}$;
\begin{align}
S_{2} \, m \, S_{2} = m \, , ~~~ 
S_{2} = 1 - 2 \bm v_{2} \otimes \bm v_{2}^{T} = 
{1\over 3}
\begin{pmatrix}
1 & -2 & -2 \\
- 2 & 1 & -2 \\
- 2 & -2 & 1 \\
\end{pmatrix} .
\label{S2}
\end{align}
This symmetry is called the {\it magic} symmetry  
and the matrix $m$ is called {\it magic} in which the row sums and the column sums are all equal to a number $\a$ \cite{Lam:2006wy}. 
The eigenvalues of $S_{2}$ are 
\begin{align}
S_{2} \bm v_{2} = - \bm v_{2} \, , ~~ S_{2} \bm  v_{1,3} = + \bm v_{1,3} \, .
\end{align}
From this, the symmetry conditions are
\begin{align}
M_{22}^{2} \, \tilde {a}_{2}^{\rm TBM} \tilde a_{1,3}^{\rm TBM} 
 = - \det M \, b_{2}^{\rm TBM} b_{1,3}^{\rm TBM}  \, . 
 \label{symcondTM2}
\end{align}

In the TBM basis, the mass matrix $m$ with these conditions is
\begin{align}
m^{\rm TBM} &\equiv  U^{T}_{\rm TBM} m \, U_{\rm TBM}  \\
& = { 1 \over \tilde M_{1}}
\begin{pmatrix}
( \tilde a_{1}^{\rm TBM})^{2} & 0 & \tilde a_{1}^{\rm TBM}  \tilde a_{3}^{\rm TBM} \\
 0 &  (\tilde a_{2}^{\rm TBM})^{2} &  0 \\
 \tilde a_{1}^{\rm TBM}  \tilde a_{3}^{\rm TBM} & 0 & (\tilde a_{3}^{\rm TBM} )^{2}
\end{pmatrix}
+  { 1 \over \tilde M_{2}}
\begin{pmatrix}
( b_{1}^{\rm TBM})^{2} & 0 & b_{1}^{\rm TBM}  b_{3}^{\rm TBM} \\
 0 & ( b_{2}^{\rm TBM})^{2} & 0 \\
b_{1}^{\rm TBM}  b_{3}^{\rm TBM} & 0 & (b_{3}^{\rm TBM})^{2} 
\end{pmatrix} \, .
\end{align}
This is indeed symmetric under $S_{2}^{\rm TBM} = {\rm diag} (1 \, , -1 \, , 1)$ in this basis. 
From this, $\th_{13}$ and $m_{2}$ are obtained as 
\begin{align}
|\sin \th_{13}^{\rm TM_{2}}| = \sqrt 2 \sin \th_{\rm TM_{2}} / \sqrt 3 \, , ~
\tan 2 \th_{\rm TM_{2}} = 
\abs{ {2  \tilde a_{1}^{\rm TBM}  \tilde a_{3}^{\rm TBM} \over \tilde M_{1}}
+ {2 b_{1}^{\rm TBM}  b_{3}^{\rm TBM} \over \tilde M_{2}}
\over 
 \abs{ {(\tilde a_{3}^{\rm TBM} )^{2} \over \tilde M_{1}} + {(b_{3}^{\rm TBM})^{2} \over \tilde M_{2}} } -
   \abs{ {(\tilde a_{1}^{\rm TBM} )^{2} \over \tilde M_{1}} + { (b_{1}^{\rm TBM})^{2} \over \tilde M_{2}}  } } \, , 
\end{align}
\begin{align}
m_{2}  & = 
{M_{22} \over \det M} ( \tilde a_{2}^{\rm TBM} )^{2} + {1\over M_{22} } ( b_{2}^{\rm TBM} )^{2} 
= 
{M_{22} \over 3 \det M} (\tilde a_{1} + \tilde a_{2} + \tilde a_{3})^{2} + {1\over 3 M_{22} } (b_{1} + b_{2} + b_{3})^{2} \, .
\end{align}

In the case of the magic symmetry, there are three types of solutions as well. 
The first four trivial solutions are
\begin{align}
S_{2} \tilde Y = \pm \tilde Y ~ \To ~ 
\tilde Y & =
 (x_{1} \bm v_{1} + x_{3} \bm v_{3} , y_{1} \bm v_{1} + y_{3} \bm v_{3} )  
 ~{\rm or}~
 (x_{2} \bm v_{2}, y_{2} \bm v_{2})
\\ & = 
\begin{pmatrix}
- \tilde a_{2} - \tilde a_{3} & - b_{2} -b_{3} \\
\tilde a_{2} & b_{2}\\
\tilde a_{3} & b_{3} \\
\end{pmatrix}
~ {\rm or} ~
\begin{pmatrix}
\tilde a_{1} & b_{1} \\
\tilde a_{1} & b_{1}  \\
\tilde a_{1} & b_{1} \\
\end{pmatrix}
\, , \\
S_{2} \tilde Y = \pm \tilde Y \s_{3} ~ \To ~ 
\tilde Y & = (x_{1} \bm v_{1} + x_{3} \bm v_{3}, y_{2} \bm v_{2} )  ~ {\rm or}~ (x_{2} \bm v_{2}, y_{1} \bm v_{1} + y_{3} \bm v_{3})  \\
& =
\begin{pmatrix}
- \tilde a_{2} - \tilde a_{3} & b_{1} \\
\tilde a_{2} & b_{1} \\
\tilde a_{3} & b_{1} \\
\end{pmatrix}
~ {\rm or} ~
\begin{pmatrix}
\tilde a_{1} & - b_{2} -b_{3} \\
\tilde a_{1} & b_{2} \\
\tilde a_{1} & b_{3} \\
\end{pmatrix} \, . 
\end{align}
The solutions with $S_{2} \tilde Y = \pm \tilde Y$ are phenomenologically excluded because they predict $m_{2} = 0$ or $m_{1,3} = 0$.
In other words, $Y$ cannot have magic (anti-)symmetry in this meaning.
In the case of $S_{2} \tilde Y = \pm \tilde Y \s_{3}$, $x_{1}, y_{1} = 0$ ($x_{3}, y_{3} = 0$) leads to a NH (IH) solution. 

The mass singular values of the trivial solutions are
\begin{align}
m_{i} & = \{ 0\, , {M_{22} \over \det M}  (\abs{\tilde a_{2}}^{2} + \abs{\tilde a_{2} + \tilde a_{3}}^{2} + \abs{\tilde a_{3}}^{2}) \, , \abs {{1\over M_{22}} 3 b_{1}^{2} } \}  \\
~~ {\rm or} ~~ & = 
\{ 0\, , \abs {{M_{22} \over \det M} 3 \tilde a_{1}^{2}} \, , {1\over M_{22}} (\abs{ b_{2}}^{2} + \abs{b_{2} + b_{3}}^{2} + \abs{b_{3}}^{2}) \} \, . 
\end{align}
Furthermore, if we impose the condition $\tilde a_{2} = \pm \tilde a_{3}$ (or $b_{2} = \pm b_{3}$),  $m$ has the $\m - \t$ symmetry and the respective signs correspond to IH and NH.

Finally, the non-trivial solution~(\ref{solution}) is analysed.
Since $m_{2}$ cannot be zero, we can set $\sqrt{3} \, \tilde a_{2}^{\rm TBM} = \sum_{i} \tilde a_{i} \neq 0$ and $\sqrt{3} \, b_{2}^{\rm TBM} = \sum_{i} b_{i} \neq 0$.
From the expansions~(\ref{expand1}) and (\ref{expand2}), 
the parameter $r$~(\ref{r}) is determined as 
\begin{align}
r_{2} 
= - {\sqrt{\det M} \over M_{22} } { b_{2}^{\rm TBM} \over \tilde a_{2}^{\rm TBM} } \, .
\end{align}
From the symmetry conditions~(\ref{symcondTM2}), $\tilde a_{1,3}$ are determined. 
Explicitly, 
\begin{align}
 {2 \tilde a_{1} - \tilde a_{2} - \tilde a_{3} } &= 
- 
{\det M \over M_{22}^{2}} {\sum_{i} b_{i} \over \sum_{i} \tilde a_{i} } 
({2 b_{1} - b_{2} - b_{3}}) \, ,  \\
{ \tilde a_{2} - \tilde a_{3} } & = 
- 
{\det M \over M_{22}^{2}} {\sum_{i} b_{i} \over \sum_{i} \tilde a_{i} } 
({b_{2} -  b_{3}})  \, , 
\end{align}
and the mass values are 
\begin{align}
m_{i} &= \{ 0\, ,  \abs{1+r_{2}^{2} \over M_{22}} ( \abs{b_{1}^{\rm TBM}}^{2} + \abs{b_{3}^{\rm TBM}}^{2} )  \, , \abs {1 + {1\over r_{2}^{2}} \over M_{22}} \abs{b_{2}^{\rm TBM} }^{2} \} \\
&= \{ 0\, ,  \abs {1+r_{2}^{2} \over M_{22}} ( { \abs{2 b_{1} - b_{2} - b_{3}}^{2} \over 6} + {\abs {b_{2} -  b_{3}}^{2} \over 2})   \, , \abs{ {1 + {1\over r_{2}^{2}} \over M_{22}} } { \abs{\sum_{i} b_{i}}^{2} \over 3} \}  \, . 
\end{align}
If we further impose the condition $b_{2} + b_{3} = 2 b_{1}$ ($b_{2} = b_{3}$), $m$ has the $\m - \t$ symmetry, corresponding to a  NH (IH) solution.

\section{Summary}

In this paper, using a formula for the minimal type-I seesaw mechanism by $LDL^{T}$ (or generalized Cholesky) decomposition, 
conditions of general $Z_{2}$-invariance of the neutrino mass matrix $m$ is obtained in an arbitrary basis.
The conditions are found to be 
$(M_{22} a_{i}^{+} - M_{12} b_{i}^{+}) \,  ( M_{22} a_{j}^{-} - M_{12} b_{j}^{-}) = 
- \det M \, b_{i}^{+} \, b_{j}^{-}$ for the $Z_{2}$-symmetric and -antisymmetric part of a Yukawa matrix $Y_{ij}^{\pm} \equiv (Y \pm T Y )_{ij} /2 \equiv  (a_{j}^{\pm}, b_{j}^{\pm})$ and the right-handed neutrino mass matrix $M_{ij}$.
In other words, the symmetric and antisymmetric part of $b_{i}$ must be proportional to those of the quantity $\tilde a_{i} \equiv a_{i} - {M_{12} \over M_{22}} b_{i}$. 
They are equivalent to the condition that $m$ is block diagonalized by eigenvectors $\bm e_{i}$ of the generator $T$. 

Since $\bm e_{i}$ are orthogonal, 
we can analyze the eigensystem of $m \, m^{\dg}$ for a $Z_{2}$-symmetric $m$.
Two eigenvectors $\bm u_{1,2}$ of $m \, m^{\dg}$ coincide with any of those of $T$, 
and the remaining one is a vector $(\bm u_{1} \times \bm u_{2})^{*}$ orthogonal to them.
Furthermore, if the Yukawa matrix does not have the $Z_{2}$ symmetry, 
two nonzero neutrino masses are represented without a radical symbol.
On the other hand, in the case of (anti-)symmetric $Y$ with $TY=\pm Y$, 
the solution of $TY= (\det T) Y$ is phenomenologically rejected because the rank of $m$ is one. 
In the other solution,  
the mass singular values cannot be expressed without solving a complicated quadratic equation.
However, if the other $Z_{2}$ symmetry can be identified, the mass values can be concisely displayed.

These results are applied to three $Z_{2}$ symmetries, the $\m-\t$ symmetry, the TM$_{1}$ mixing, and the magic symmetry which predicts the TM$_{2}$ mixing. 
For the case of TM$_{1,2}$, the symmetry conditions becomes 
$ M_{22}^{2} \, \tilde {a}_{1}^{\rm TBM} \tilde a_{2}^{\rm TBM} 
 = - \det M \, b_{1}^{\rm TBM}  b_{2}^{\rm TBM}  $ 
 and 
 $ M_{22}^{2} \, \tilde {a}_{1,2}^{\rm TBM} \tilde a_{3}^{\rm TBM} 
 = - \det M \, b_{1,2}^{\rm TBM} b_{3}^{\rm TBM}$
with components $\tilde a_{i}^{\rm TBM}$ and $b_{i}^{\rm TBM}$ in the TBM basis $\bm v_{1,2,3}$.
In particular, for the TM$_{2}$ mixing, the magic (anti-)symmetric Yukawa matrix with $S_{2} Y = \pm Y$ 
 is phenomenologically excluded because it predicts $m_{2}=0$ or $m_{1}, m_{3} = 0$.

\section*{Acknowledgement}

This study is financially supported 
by JSPS Grants-in-Aid for Scientific Research
No.~18H01210 and MEXT KAKENHI Grant No.~18H05543.

\appendix

\section{Understanding from the original raw formula}

Let us consider the condition~(\ref{cond3}) without the $LDL^{T}$ decomposition.
By 
rewriting $\tilde {\bm a}$ to $\bm a$, 
\begin{align}
(M_{22} a_{i}^{+} - M_{12} b_{i}^{+}) \,  ( M_{22} a_{j}^{-} - M_{12} b_{j}^{-}) = 
- \det M \, b_{i}^{+} \, b_{j}^{-} \,  .
\label{cond4}
\end{align}
By adding this equation with $i$ and $j$ interchanged, 
\begin{align}
{\rm Asym} [ (a_{i} M_{22} - b_{i} M_{12}) a_{j} ] &= 
 - {\rm Asym} [  ( b_{i} M_{11}  - a_{i} M_{12}) b_{j}] \, , 
\label{cond5}
\end{align}
where ${\rm Asym} [x_{i} y_{j}] \equiv x_{i}^{+} y_{j}^{-} + x_{i}^{-} y_{j}^{+}$
is the antisymmetric part for $T$. 

They are identical to the conditions $T m T^{T} = m$ in the original raw formula.
From Eq.~(\ref{1}), the mass matrix $m$ is written by 
\begin{align}
m &= Y M^{-1} Y^{T} = 
{1 \over \det M} 
\begin{pmatrix}
a_{1} & b_{1} \\
a_{2} & b_{2} \\
a_{3} & b_{3} \\
\end{pmatrix}
\begin{pmatrix}
- (\bm M_{2} \times \bm Y_{1})_{z} & - (\bm M_{2} \times \bm Y_{2})_{z} & -(\bm M_{2} \times \bm Y_{3})_{z} \\
(\bm M_{1} \times \bm Y_{1})_{z} & (\bm M_{1} \times \bm Y_{2})_{z} & (\bm M_{1} \times \bm Y_{3})_{z} \\
\end{pmatrix} \, , \\
m_{ij} & = {1\over \det M} [a_{i} (a_{j} M_{22} -  b_{j} M_{21}) + b_{i}  ( b_{j} M_{11} -  a_{j} M_{12})] \, , 
\end{align}
where $(\bm M_{i} \times \bm Y_{j})_{z} \equiv M_{i1} Y_{j2} - M_{i2} Y_{j1}$.  
Since the asymmetric parts of the two matrices with rank one must cancel for $T$-invariant $m$,  Eq.~(\ref{cond5}) is obtained as the condition.
Also, by considering these conditions from symmetries, 
the solution can be displayed by a complex orthogonal matrix $O$. 
This point is discussed in the next section.

\section{Understanding from complex orthogonal matrices} 

The solutions~(\ref{solution}) and (\ref{solution2}) can also be understood from orthgonal matrices.
By defining $X \equiv \tilde Y \sqrt{\tilde M^{-1}}$, the mass matrix $m$ is written only in $X$; 
\begin{align}
X = \row{ \sqrt{\dfrac{M_{22}}{\det M}} \tilde {\bm a}}{\sqrt{\dfrac{1}{M_{22}}}\bm b} \, , ~~~
m = X X^{T} \, .
\end{align}
In order for $m$ to have $T$ symmetry, 
$X = (u , v)$ must satisfy the following transformation with a complex orthogonal matrix $O$; 
\begin{align}
T X =  (T u \, , \, T v) = \pm X O \, .
\label{symmetry}
\end{align}
Since $T^{2} X = \pm T X O=  X O^{2} = X$ holds,  
the matrix $O$ satisfies $O = O^{-1}=O^T$. 
In other words, $O$ is a symmetric orthogonal matrix.

In the case of $\det O = +1$, only $O = \pm 1_{3}$ is allowed since $\sin z = 0$.
This corresponds to $T$-(anti)symmetric $\tilde Y$ and $Y$ in Eq.~(\ref{solution2}), respectively.
On the other hand, in the case of $\det O = -1$, 
\begin{align}
 O = 
\begin{pmatrix}
 \cos z & \sin z \\
 \sin z & -\cos z \\
\end{pmatrix} \, , 
\label{O}
\end{align}
and there is no restriction on the complex parameter $z$.
The $T$-invariant condition becomes 
\begin{align}
X O = (u , v) 
\begin{pmatrix}
\cos z & \sin z \\
\sin z & - \cos z \\
\end{pmatrix}
= ( c_{z} u +  s_{z} v , \, s_{z} u - c_{z} v)
= \pm (T u \, , \, T v) = \pm T X \, , 
\end{align}
where $c_{z} \equiv \cos z , s_{z} \equiv \sin z$. From this, 
\begin{align}
 u = {1 \over s_{z}} (\pm T v + c_{z} v) \, ,  ~~~
T u = \pm {c_{z} \over s_{z}} (\pm T v + c_{z} v) \pm s_{z} v
= {1\over s_{z}} (c_{z} T v  \pm   v) \, .
\end{align}
From $u^{\pm} = {1\over 2} ( u \pm T u)$ and $v^{\pm} = {1\over 2} ( v \pm T v)$, we obtain
\begin{align}
 \sqrt{\dfrac{M_{22}}{\det M}} \tilde {\bm a}^{+}  = 
u^{+} & 
= { c_{z} \pm 1 \over s_{z}} v^{+}
= \{\cot {z \over 2} \, , \, -\tan {z \over 2} \} \sqrt{\dfrac{1}{M_{22}}} \bm b^{+} \, , \label{rep1} \\
 \sqrt{\dfrac{M_{22}}{\det M}} \tilde {\bm a}^{-}  = 
u^{-} & 
= { c_{z} \mp 1 \over s_{z}} v^{-}
=  \{ -\tan {z \over 2} \, , \, \cot {z \over 2} \}  \sqrt{\dfrac{1}{M_{22}}} \bm b^{-} \, ,  
\label{rep2}
\end{align}
Obviously,  these coefficients satisfy 
\begin{align}
{ c_{z} \pm 1 \over s_{z}} \times {c_{z} \mp 1 \over s_{z}} = -1,
\end{align}
and it corresponds to the solution~(\ref{solution}).
Note that the sign $\pm$ comes from the parity for $T$.

An understanding of these solutions by matrices is as follows. 
As in the previous paper \cite{Yang:2022wch}, we consider a quantity $X(1 \pm O \s_{3})$.
This extra $\s_{3}$ has an effect of exchanging $\bm b^{+}$ and $\bm b^{-}$. For example, in the case of $XO = + TX$, 
\begin{align}
X (1 + O \s_{3}) & = X + T X \s_{3} = 2 ( \tilde {\bm a}^{+} \, ,  \bm b^{-})  \sqrt{\tilde M ^{-1}} \, ,  \label{xpo} \\
X (1 - O \s_{3}) &  = X - T X \s_{3} = 2 ( \tilde {\bm a}^{-} \, , \bm b^{+}) \sqrt{\tilde M ^{-1} } \, . \label{xmo}
\end{align}
Since $O \s_{3}$ is written as
\begin{align}
 O \s_{3}
 =
\begin{pmatrix}
c_{z} & - s_{z} \\
s_{z} & c_{z}
\end{pmatrix}
= 
c_{z} 1_{2} - i \s_{2} s_{z} \, , 
\end{align}
the term is expressed as $1 \pm O \s_{3} = 1_{2} (1 \pm c_{z})  \mp i \s_{2} s_{z} $. 
From this, 
\begin{align}
X (1 \pm O \s_{3}) {1 \mp c_{z} \over s_{z}} & = X [1_{2} s_{z} \mp  i \s_{2} (1 \mp c_{z}) ]
= \mp X (1 \mp O \s_{3}) i \s_{2} \, . 
\end{align}
For example, matrix elements of the upper sign become
\begin{align}
(  \tilde {\bm a}^{+} \, ,  \bm b^{-}) \sqrt{\tilde M^{-1}} 
= {- s_{z} \over 1 - c_{z}}  (\tilde {\bm a}^{-} \, , \bm b^{+}) \sqrt{\tilde M^{-1}}
 \offdiag{1}{-1} 
= {s_{z} \over 1 - c_{z}}  (\sqrt{\frac{1}{M_{22}}} \bm b^{+} \, , - \sqrt{\frac{M_{22}}{\det M}}  \tilde {\bm a}^{-})
\, , 
\end{align}
and they correspond to Eqs.~(\ref{rep1}) and (\ref{rep2}).

The symmetry condition~(\ref{cond3}) satisfied by these solutions, {\it i.e.} $M_{22}^{2} \, \tilde a_{i}^{+} \, \tilde a_{j}^{-} = -  |M| \, b_{i}^{+} \, b_{j}^{-}$ is obviously equivalent to $X^{+} X^{- \, T} = 0$ for $X^{\pm} \equiv (X \pm T X)/2$. 
This condition can be rewritten as
\begin{align}
X^{+} X^{- \, T} = {1\over 4}  X (1\pm O)(1 \mp O)^{T} X^{T} 
=  \pm {1\over 4i}  X (O - O^{T}) X^{T} = 0 \, , 
\end{align}
and it is a solution because $O^{T} = O$ holds.

Furthermore, a matrix representation of the solution $T X = \pm X O$ is explored.
In the case of $\det O= + 1$, $TX_{\pm} = \pm X_{\pm}$ holds and
 $X$ is symmetric or antisymmetric under $T$, as discussed above.
In the case of $\det O = -1$, we consider the following solution 
by a complex orthogonal matrix $Q$ and $\s_{3} \equiv {\rm diag} (1, -1)$.
\begin{align}
T X_{1} = X_{1} \s_{3} \, , ~~~
X_{1}' = X_{1} Q \, .
\end{align}
This $X_{1} = (u^{+} , v^{-})$ has $T$-symmetric and -antisymmetric vectors
\footnote{The other solution $X_{2} = (u^{+} , v^{-})$ can be reached by a permutation with $\det O = -1$ from $X_{1}$.}. 
The $T$ transformation for $X'_{1}$ is
\begin{align}
T X_{1}' = X_{1} \s_{3} Q = X_{1}' Q^{T} \s_{3} Q \, . 
\end{align}
Thus, if we define $Q^{T} \s_{3} Q \equiv O_{1}$, this is a symmetric orthogonal matrix with $\det O_{1} = -1$ and $X_{1}'$ satisfies the symmetry~(\ref{symmetry}).
\begin{align}
T X'_{1}= X'_{1} O_{1} \, .
\end{align}
Specifically, 
\begin{align}
Q \equiv 
\begin{pmatrix}
\cos w & - \sin w \\
 \sin w & \cos w \\
\end{pmatrix}
~~ \To ~~
O_{1} &= Q^{T} \s_{3} Q = 
\begin{pmatrix}
\cos 2 w & - \sin 2 w \\
- \sin 2 w & - \cos 2 w
\end{pmatrix} \, .
\end{align}
Indeed $O_{1}$ is symmetric, and it agrees with Eq.~(\ref{O}) by a suitable redefinition.

Since we can write $X = X_{\pm}$ or $X_{1} Q$, 
the general Yukawa matrix for a $T$-invariant $m$ can be written by these matrices. 
In the case of $X = X_{\pm}, $ Yukawa matrices $Y$ and $\tilde Y$ are 
\begin{align}
\tilde Y_{\pm} = X_{\pm} \sqrt{\tilde M}  \, , ~~~
Y_{\pm} = \tilde Y L^{-1} = X_{\pm} \sqrt{\tilde M} L^{-1} \, . 
\end{align}
In other words, $Y_{\pm}$ is symmetric or antisymmetric under $T$, $T Y_{\pm} = \pm Y$, and it can be expanded by eigenvectors of $T$ with the same eigenvalues. 

In the other case, 
\begin{align}
\tilde Y = X \sqrt{\tilde M} = X_{1} Q \sqrt{\tilde M} \, , ~~~
Y = \tilde Y L^{-1} = X_{1} Q \sqrt{\tilde M} L^{-1} \, . 
\end{align}
Since $X_{1}$ has the form $X_{1} = (u^{-} , v^{+})$, it represents the solution~(\ref{rep1}) and (\ref{rep2}). 

From the diagonalization $L^{-1} M^{-1} (L^{T})^{-1} = \tilde M^{-1}$~(\ref{tildeM}), 
we finally obtain 
\begin{align}
m = Y M^{-1} Y^{T} =  X_{\pm, 1} Q Q^{T} X_{\pm , 1}^{T} \, ,
\end{align}
and $m$ is $T$-invariant.
Therefore, in order to predict $T$-invariant $m$, 
the Yukawa matrix $Y$ must be $T$-symmetric or antisymmetric, or 
has degrees of freedom complex orthogonal matrix $Q$ multiplied to 
$T$-covariant $X_{1}$ that satisfies $T X_{1} = X_{1} \s_{3}$. 



\begin{thebibliography}{100}

\bibitem{Barger:1998ta}
V.~D. Barger, S.~Pakvasa, T.~J. Weiler, and K.~Whisnant,
\newblock Phys. Lett. B {\bf 437}, 107 (1998), arXiv:hep-ph/9806387.

\bibitem{Fukuyama:1997ky}
T.~Fukuyama and H.~Nishiura,
\newblock (1997), arXiv:hep-ph/9702253.

\bibitem{Lam:2001fb}
C.~S. Lam,
\newblock Phys. Lett. {\bf B507}, 214 (2001), arXiv:hep-ph/0104116.

\bibitem{Ma:2001mr}
E.~Ma and M.~Raidal,
\newblock Phys. Rev. Lett. {\bf 87}, 011802 (2001), arXiv:hep-ph/0102255,
\newblock [Erratum: Phys. Rev. Lett.87,159901(2001)].

\bibitem{Balaji:2001ex}
K.~R.~S. Balaji, W.~Grimus, and T.~Schwetz,
\newblock Phys. Lett. {\bf B508}, 301 (2001), arXiv:hep-ph/0104035.

\bibitem{Koide:2002cj}
Y.~Koide, H.~Nishiura, K.~Matsuda, T.~Kikuchi, and T.~Fukuyama,
\newblock Phys. Rev. {\bf D66}, 093006 (2002), arXiv:hep-ph/0209333.

\bibitem{Kitabayashi:2002jd}
T.~Kitabayashi and M.~Yasue,
\newblock Phys. Rev. {\bf D67}, 015006 (2003), arXiv:hep-ph/0209294.

\bibitem{Harrison:2002et}
P.~F. Harrison and W.~G. Scott,
\newblock Phys. Lett. {\bf B547}, 219 (2002), arXiv:hep-ph/0210197.

\bibitem{Grimus:2003yn}
W.~Grimus and L.~Lavoura,
\newblock Phys. Lett. {\bf B579}, 113 (2004), arXiv:hep-ph/0305309.

\bibitem{Koide:2003rx}
Y.~Koide,
\newblock Phys. Rev. {\bf D69}, 093001 (2004), arXiv:hep-ph/0312207.

\bibitem{Ghosal:2003mq}
A.~Ghosal,
\newblock (2003), arXiv:hep-ph/0304090.

\bibitem{Aizawa:2004qf}
I.~Aizawa, M.~Ishiguro, T.~Kitabayashi, and M.~Yasue,
\newblock Phys. Rev. {\bf D70}, 015011 (2004), arXiv:hep-ph/0405201.

\bibitem{Ghosal:2004qb}
A.~Ghosal,
\newblock Mod. Phys. Lett. {\bf A19}, 2579 (2004).

\bibitem{Koide:2004gj}
Y.~Koide,
\newblock Phys. Lett. {\bf B607}, 123 (2005), arXiv:hep-ph/0411280.

\bibitem{Mohapatra:2005yu}
R.~N. Mohapatra and W.~Rodejohann,
\newblock Phys. Rev. {\bf D72}, 053001 (2005), arXiv:hep-ph/0507312.

\bibitem{Kitabayashi:2005fc}
T.~Kitabayashi and M.~Yasue,
\newblock Phys. Lett. {\bf B621}, 133 (2005), arXiv:hep-ph/0504212.

\bibitem{Joshipura:2005vy}
A.~S. Joshipura,
\newblock Eur. Phys. J. {\bf C53}, 77 (2008), arXiv:hep-ph/0512252.

\bibitem{Haba:2006hc}
N.~Haba and W.~Rodejohann,
\newblock Phys. Rev. {\bf D74}, 017701 (2006), arXiv:hep-ph/0603206.

\bibitem{Xing:2006xa}
Z.-z. Xing, H.~Zhang, and S.~Zhou,
\newblock Phys. Lett. {\bf B641}, 189 (2006), arXiv:hep-ph/0607091.

\bibitem{Ahn:2006nu}
Y.~H. Ahn, S.~K. Kang, C.~S. Kim, and J.~Lee,
\newblock Phys. Rev. {\bf D73}, 093005 (2006), arXiv:hep-ph/0602160.

\bibitem{Joshipura:2007sf}
A.~S. Joshipura and B.~P. Kodrani,
\newblock Phys. Lett. {\bf B670}, 369 (2009), arXiv:0706.0953.

\bibitem{GomezIzquierdo:2009id}
J.~C. Gomez-Izquierdo and A.~Perez-Lorenzana,
\newblock Phys. Rev. {\bf D82}, 033008 (2010), arXiv:0912.5210.

\bibitem{Xing:2010ez}
Z.-z. Xing and Y.-L. Zhou,
\newblock Phys. Lett. {\bf B693}, 584 (2010), arXiv:1008.4906.

\bibitem{Ge:2010js}
S.-F. Ge, H.-J. He, and F.-R. Yin,
\newblock JCAP {\bf 1005}, 017 (2010), arXiv:1001.0940.

\bibitem{He:2011kn}
H.-J. He and F.-R. Yin,
\newblock Phys. Rev. D {\bf 84}, 033009 (2011), arXiv:1104.2654.

\bibitem{He:2012yt}
H.-J. He and X.-J. Xu,
\newblock Phys. Rev. D {\bf 86}, 111301 (2012), arXiv:1203.2908.

\bibitem{Xing:2015fdg}
Z.-z. Xing and Z.-h. Zhao,
\newblock Rept. Prog. Phys. {\bf 79}, 076201 (2016), arXiv:1512.04207.

\bibitem{He:2015afa}
X.-G. He,
\newblock Chin. J. Phys. {\bf 53}, 100101 (2015), arXiv:1504.01560.

\bibitem{He:2015xha}
H.-J. He, W.~Rodejohann, and X.-J. Xu,
\newblock Phys. Lett. {\bf B751}, 586 (2015), arXiv:1507.03541.

\bibitem{Xing:2017cwb}
Z.-z. Xing and J.-y. Zhu,
\newblock Chin. Phys. {\bf C41}, 123103 (2017), arXiv:1707.03676.

\bibitem{Gomez-Izquierdo:2017rxi}
J.~C. G{\'o}mez-Izquierdo,
\newblock Eur. Phys. J. {\bf C77}, 551 (2017), arXiv:1701.01747.

\bibitem{Fukuyama:2017qxb}
T.~Fukuyama,
\newblock PTEP {\bf 2017}, 033B11 (2017), arXiv:1701.04985.

\bibitem{Harrison:2002kp}
P.~F. Harrison and W.~G. Scott,
\newblock Phys. Lett. {\bf B535}, 163 (2002), arXiv:hep-ph/0203209.

\bibitem{Friedberg:2006it}
R.~Friedberg and T.~D. Lee,
\newblock HEPNP {\bf 30}, 591 (2006), arXiv:hep-ph/0606071.

\bibitem{Bjorken:2005rm}
J.~D. Bjorken, P.~F. Harrison, and W.~G. Scott,
\newblock Phys. Rev. {\bf D74}, 073012 (2006), arXiv:hep-ph/0511201.

\bibitem{He:2006qd}
X.-G. He and A.~Zee,
\newblock Phys. Lett. {\bf B645}, 427 (2007), arXiv:hep-ph/0607163.

\bibitem{Grimus:2008tt}
W.~Grimus and L.~Lavoura,
\newblock JHEP {\bf 09}, 106 (2008), arXiv:0809.0226.

\bibitem{Channey:2018cfj}
K.~S. Channey and S.~Kumar,
\newblock J. Phys. G {\bf 46}, 015001 (2019), arXiv:1812.10268.

\bibitem{Bao:2021zwu}
H.-C. Bao, X.-Y. Zhao, and Z.-H. Zhao,
\newblock (2021), arXiv:2104.05394.

\bibitem{Lam:2006wy}
C.~S. Lam,
\newblock Phys. Lett. {\bf B640}, 260 (2006), arXiv:hep-ph/0606220.

\bibitem{Harrison:2002er}
P.~F. Harrison, D.~H. Perkins, and W.~G. Scott,
\newblock Phys. Lett. {\bf B530}, 167 (2002), arXiv:hep-ph/0202074.

\bibitem{Lam:2006wm}
C.~S. Lam,
\newblock Phys. Rev. D {\bf 74}, 113004 (2006), arXiv:hep-ph/0611017.

\bibitem{Lam:2007qc}
C.~S. Lam,
\newblock Phys. Lett. B {\bf 656}, 193 (2007), arXiv:0708.3665.

\bibitem{Lam:2008rs}
C.~S. Lam,
\newblock Phys. Rev. Lett. {\bf 101}, 121602 (2008), arXiv:0804.2622.

\bibitem{Ma:1998zg}
E.~Ma, D.~P. Roy, and U.~Sarkar,
\newblock Phys. Lett. B {\bf 444}, 391 (1998), arXiv:hep-ph/9810309.

\bibitem{King:1998jw}
S.~F. King,
\newblock Phys. Lett. B {\bf 439}, 350 (1998), arXiv:hep-ph/9806440.

\bibitem{Frampton:2002qc}
P.~H. Frampton, S.~L. Glashow, and T.~Yanagida,
\newblock Phys. Lett. B {\bf 548}, 119 (2002), arXiv:hep-ph/0208157.

\bibitem{Xing:2020ald}
Z.-z. Xing and Z.-h. Zhao,
\newblock Rept. Prog. Phys. {\bf 84}, 066201 (2021), arXiv:2008.12090.

\bibitem{Guo:2003cc}
W.-l. Guo and Z.-z. Xing,
\newblock Phys. Lett. B {\bf 583}, 163 (2004), arXiv:hep-ph/0310326.

\bibitem{Barger:2003gt}
V.~Barger, D.~A. Dicus, H.-J. He, and T.-j. Li,
\newblock Phys. Lett. B {\bf 583}, 173 (2004), arXiv:hep-ph/0310278.

\bibitem{Mei:2003gn}
J.-w. Mei and Z.-z. Xing,
\newblock Phys. Rev. D {\bf 69}, 073003 (2004), arXiv:hep-ph/0312167.

\bibitem{Chang:2004wy}
S.~Chang, S.~K. Kang, and K.~Siyeon,
\newblock Phys. Lett. B {\bf 597}, 78 (2004), arXiv:hep-ph/0404187.

\bibitem{Guo:2006qa}
W.-l. Guo, Z.-z. Xing, and S.~Zhou,
\newblock Int. J. Mod. Phys. E {\bf 16}, 1 (2007), arXiv:hep-ph/0612033.

\bibitem{Kitabayashi:2007bs}
T.~Kitabayashi,
\newblock Phys. Rev. D {\bf 76}, 033002 (2007), arXiv:hep-ph/0703303.

\bibitem{He:2009pt}
X.-G. He and W.~Liao,
\newblock Phys. Lett. B {\bf 681}, 253 (2009), arXiv:0909.1463.

\bibitem{Yang:2011fh}
R.-Z. Yang and H.~Zhang,
\newblock Phys. Lett. B {\bf 700}, 316 (2011), arXiv:1104.0380.

\bibitem{Harigaya:2012bw}
K.~Harigaya, M.~Ibe, and T.~T. Yanagida,
\newblock Phys. Rev. D {\bf 86}, 013002 (2012), arXiv:1205.2198.

\bibitem{Kitabayashi:2016zec}
T.~Kitabayashi and M.~Yasu\`e,
\newblock Phys. Rev. D {\bf 94}, 075020 (2016), arXiv:1605.04402.

\bibitem{Bambhaniya:2016rbb}
G.~Bambhaniya, P.~S. Bhupal~Dev, S.~Goswami, S.~Khan, and W.~Rodejohann,
\newblock Phys. Rev. D {\bf 95}, 095016 (2017), arXiv:1611.03827.

\bibitem{Li:2017zmk}
C.-C. Li and G.-J. Ding,
\newblock Phys. Rev. D {\bf 96}, 075005 (2017), arXiv:1701.08508.

\bibitem{Liu:2017frs}
Z.-C. Liu, C.-X. Yue, and Z.-h. Zhao,
\newblock JHEP {\bf 10}, 102 (2017), arXiv:1707.05535.

\bibitem{Shimizu:2017fgu}
Y.~Shimizu, K.~Takagi, and M.~Tanimoto,
\newblock JHEP {\bf 11}, 201 (2017), arXiv:1709.02136.

\bibitem{Shimizu:2017vwi}
Y.~Shimizu, K.~Takagi, and M.~Tanimoto,
\newblock Phys. Lett. B {\bf 778}, 6 (2018), arXiv:1711.03863.

\bibitem{Nath:2018hjx}
N.~Nath, Z.-z. Xing, and J.~Zhang,
\newblock Eur. Phys. J. {\bf C78}, 289 (2018), arXiv:1801.09931.

\bibitem{Barreiros:2018bju}
D.~M. Barreiros, R.~G. Felipe, and F.~R. Joaquim,
\newblock JHEP {\bf 01}, 223 (2019), arXiv:1810.05454.

\bibitem{Nath:2018xih}
N.~Nath,
\newblock Mod. Phys. Lett. A {\bf 34}, 1950329 (2019), arXiv:1808.05062.

\bibitem{Wang:2019ovr}
X.~Wang and S.~Zhou,
\newblock JHEP {\bf 05}, 017 (2020), arXiv:1910.09473.

\bibitem{Zhao:2020bzx}
Z.-h. Zhao,
\newblock JHEP {\bf 11}, 170 (2021), arXiv:2003.00654.

\bibitem{Ecker:1981wv}
G.~Ecker, W.~Grimus, and W.~Konetschny,
\newblock Nucl. Phys. B {\bf 191}, 465 (1981).

\bibitem{Ecker:1983hz}
G.~Ecker, W.~Grimus, and H.~Neufeld,
\newblock Nucl. Phys. B {\bf 247}, 70 (1984).

\bibitem{Gronau:1985sp}
M.~Gronau and R.~N. Mohapatra,
\newblock Phys. Lett. B {\bf 168}, 248 (1986).

\bibitem{Ecker:1987qp}
G.~Ecker, W.~Grimus, and H.~Neufeld,
\newblock J. Phys. A {\bf 20}, L807 (1987).

\bibitem{Neufeld:1987wa}
H.~Neufeld, W.~Grimus, and G.~Ecker,
\newblock Int. J. Mod. Phys. A {\bf 3}, 603 (1988).

\bibitem{Ferreira:2009wh}
P.~Ferreira, H.~E. Haber, and J.~P. Silva,
\newblock Phys. Rev. D {\bf 79}, 116004 (2009), arXiv:0902.1537.

\bibitem{Feruglio:2012cw}
F.~Feruglio, C.~Hagedorn, and R.~Ziegler,
\newblock JHEP {\bf 07}, 027 (2013), arXiv:1211.5560.

\bibitem{Holthausen:2012dk}
M.~Holthausen, M.~Lindner, and M.~A. Schmidt,
\newblock JHEP {\bf 04}, 122 (2013), arXiv:1211.6953.

\bibitem{Ding:2013bpa}
G.-J. Ding, S.~F. King, and A.~J. Stuart,
\newblock JHEP {\bf 12}, 006 (2013), arXiv:1307.4212.

\bibitem{Girardi:2013sza}
I.~Girardi, A.~Meroni, S.~Petcov, and M.~Spinrath,
\newblock JHEP {\bf 02}, 050 (2014), arXiv:1312.1966.

\bibitem{Nishi:2013jqa}
C.~Nishi,
\newblock Phys. Rev. D {\bf 88}, 033010 (2013), arXiv:1306.0877.

\bibitem{Ding:2013hpa}
G.-J. Ding, S.~F. King, C.~Luhn, and A.~J. Stuart,
\newblock JHEP {\bf 05}, 084 (2013), arXiv:1303.6180.

\bibitem{Feruglio:2013hia}
F.~Feruglio, C.~Hagedorn, and R.~Ziegler,
\newblock Eur. Phys. J. C {\bf 74}, 2753 (2014), arXiv:1303.7178.

\bibitem{Chen:2014wxa}
P.~Chen, C.-C. Li, and G.-J. Ding,
\newblock Phys. Rev. D {\bf 91}, 033003 (2015), arXiv:1412.8352.

\bibitem{Ding:2014ora}
G.-J. Ding, S.~F. King, and T.~Neder,
\newblock JHEP {\bf 12}, 007 (2014), arXiv:1409.8005.

\bibitem{Ding:2014hva}
G.-J. Ding and Y.-L. Zhou,
\newblock JHEP {\bf 06}, 023 (2014), arXiv:1404.0592.

\bibitem{Chen:2014tpa}
M.-C. Chen, M.~Fallbacher, K.~Mahanthappa, M.~Ratz, and A.~Trautner,
\newblock Nucl. Phys. B {\bf 883}, 267 (2014), arXiv:1402.0507.

\bibitem{Li:2015jxa}
C.-C. Li and G.-J. Ding,
\newblock JHEP {\bf 05}, 100 (2015), arXiv:1503.03711.

\bibitem{Turner:2015uta}
J.~Turner,
\newblock Phys. Rev. D {\bf 92}, 116007 (2015), arXiv:1507.06224.

\bibitem{Rodejohann:2017lre}
W.~Rodejohann and X.-J. Xu,
\newblock Phys. Rev. {\bf D96}, 055039 (2017), arXiv:1705.02027.

\bibitem{Penedo:2017vtf}
J.~Penedo, S.~Petcov, and A.~Titov,
\newblock JHEP {\bf 12}, 022 (2017), arXiv:1705.00309.

\bibitem{Nath:2018fvw}
N.~Nath, R.~Srivastava, and J.~W. Valle,
\newblock Phys. Rev. D {\bf 99}, 075005 (2019), arXiv:1811.07040.

\bibitem{Yang:2020qsa}
M.~J.~S. Yang,
\newblock Phys. Lett. B {\bf 806}, 135483 (2020), arXiv:2002.09152.

\bibitem{Yang:2020goc}
M.~J.~S. Yang,
\newblock Chin. Phys. C {\bf 45}, 043103 (2021), arXiv:2003.11701.

\bibitem{Yang:2021smh}
M.~J.~S. Yang,
\newblock Nucl. Phys. B {\bf 972}, 115549 (2021), arXiv:2103.12289.

\bibitem{Yang:2021xob}
M.~J.~S. Yang,
\newblock PTEP {\bf 2022}, 013B12 (2021), arXiv:2104.12063.

\bibitem{Yang:2021arl}
M.~J.~S. Yang,
\newblock PTEP {\bf 2022}, 051B01, arXiv:2112.14402.

\bibitem{Yang:2022wch}
M.~J.~S. Yang,
\newblock (2022), arXiv:2202.10067.

\bibitem{Yang:2021byq}
M.~J.~S. Yang,
\newblock PTEP {\bf 2022}, 043B05, arXiv:2110.10907.

\bibitem{Fujihara:2005pv}
T.~Fujihara {\em et~al.},
\newblock Phys. Rev. D {\bf 72}, 016006 (2005), arXiv:hep-ph/0505076.

\bibitem{Albright:2008rp}
C.~H. Albright and W.~Rodejohann,
\newblock Eur. Phys. J. {\bf C62}, 599 (2009), arXiv:0812.0436.

\bibitem{Albright:2010ap}
C.~H. Albright, A.~Dueck, and W.~Rodejohann,
\newblock Eur. Phys. J. {\bf C70}, 1099 (2010), arXiv:1004.2798.

\bibitem{He:2011gb}
X.-G. He and A.~Zee,
\newblock Phys. Rev. D {\bf 84}, 053004 (2011), arXiv:1106.4359.

\bibitem{Luhn:2013lkn}
C.~Luhn,
\newblock Nucl. Phys. B {\bf 875}, 80 (2013), arXiv:1306.2358.

\bibitem{Li:2013jya}
C.-C. Li and G.-J. Ding,
\newblock Nucl. Phys. B {\bf 881}, 206 (2014), arXiv:1312.4401.

\bibitem{King:2019vhv}
S.~F. King and Y.-L. Zhou,
\newblock Phys. Rev. D {\bf 101}, 015001 (2020), arXiv:1908.02770.

\bibitem{Krishnan:2020xeq}
R.~Krishnan, A.~Mukherjee, and S.~Goswami,
\newblock JHEP {\bf 09}, 050 (2020), arXiv:2001.07388.

\end{thebibliography}

\end{document}